# Collinear features impair visual detection by rats

Philip Meier*, Erik Flister*, Pamela Reinagel

## Abstract

We measure rats' ability to detect an oriented visual target grating located between two flanking stimuli ("flankers"). Flankers varied in contrast, orientation, angular position, and sign. Rats are impaired at detecting visual targets with collinear flankers, compared to configurations where flankers differ from the target in orientation or angular position. In particular, rats are more likely to miss the target when flankers are collinear. The same impairment is found even when the flanker luminance was sign-reversed relative to the target. These findings suggest that contour alignment alters visual processing in rats, despite their lack of orientation columns in visual cortex. This is the first report that the arrangement of visual features relative to each other affects visual behavior in rats. To provide a conceptual framework for our findings, we relate our stimuli to a contrast normalization model of early visual processing. We suggest a pattern-sensitive generalization of the model which could account for a collinear deficit. These experiments were performed using a novel method for automated high-throughput training and testing of visual behavior in rodents.

## Introduction

In this study, we consider a classic task of visual psychophysics, the discrimination between the presence and absence of a visual target at a known location. Human perception of oriented targets is influenced by the contrast, spatial frequency and orientation of nearby stimuli, both in contrast discrimination tasks(Ejima and Takahashi, 1985; Xing and Heeger, 2001) and target detection tasks(Chen and Tyler, 2008; Polat and Sagi, 1993, 2007; Solomon and Morgan, 2000; Williams and Hess, 1998; Zenger and Sagi, 1996). This influence is typically largest when the surrounding stimuli match the orientation and spatial frequency of the target, both for annuli that completely surround a target and for discrete flankers(Cannon and Fullenkamp, 1996; Chubb et al., 1989; Polat and Sagi, 1993). The influence of stimulus arrangement and phase are more variable and subject to experimental paradigms. The behavioral influence of flanking stimuli has occasionally been studied in non-human primates(Li et al., 2006), but never in rodents.

The psychophysical effect of flankers may be caused by surround processing in visual neurons, whereby features outside of the classical receptive field modulate neural responses(Angelucci et al., 2002; Chisum and Fitzpatrick, 2004). The amplitude of neural responses in the retina, thalamus, and visual cortex are normalized to spatially nearby contrast(Carandini et al., 1997; Heeger, 1992; Shapley and Victor, 1979), likely due to lateral connectivity at each level. In many cases oriented stimuli in the surround suppress activity, and suppress most when surround orientation matches the driving stimulus(Bonds, 1989; Cavanaugh et al., 2002; Polat et al., 1998). However, some cells in the appropriate contrast conditions increase their spiking activity when the orientation of stimuli in the surround matches the driving stimulus(Li et al., 2006; Polat et al., 1998; Sillito et al., 1993). In many physiology experiments, oriented surround stimuli are presented in an annulus. However, physiological surround effects can depend on the angular position of the flanker with respect to the target orientation, specifically influencing geometric arrangements like collinearity(Cavanaugh et al., 2002; Polat et al., 1998). Cortical circuits are likely to be involved in orientation-specific surround processing(Chisum et al., 2003; Das and Gilbert, 1999; Gilbert and Wiesel, 1989). It remains unknown if the orientation-selective circuits described in cats and primates are also found in rats. Rodents have orientation tuned cells in V1 but lack orientation columns(Ohki et al., 2005; Van Hooser et al., 2006).

From a theoretical perspective there are several reasons it would be advantageous for representations of local features to be sensitive to nearby image context(Series et al., 2003). In natural scenes, local image features such as luminance, contrast, and orientation are correlated at nearby locations(Field, 1987; Geisler, 2008; Ruderman and Bialek, 1994). When features are spatially correlated, surround processing can optimize the fidelity or efficiency of image estimation(Barlow, 2001; Geisler, 2008). For example divisive normalization from a nearby population of cells(Heeger, 1992) can allow a neuron to better adapt its sensitivity, reduce redundancy with its neighbors, and thus maximize information transfer(Schwartz and Simoncelli, 2001). Surround processing could also enhance salience of relevant features such as continuous contours(Field et al., 1993; Geisler et al., 2001; Sigman et al., 2001) or statistically surprising features(Itti and Koch, 2000). These theories and others predict that different patterns in the surround should have distinct influences on a visual target's neuronal representation, even if lower order statistics like luminance and contrast are matched.

In the interest of developing a rodent model for the study of surround processing, we trained rats to report the presence or absence of an oriented target when sandwiched between two flanking stimuli. Rats have previously been trained on visual tasks including grating detection(Birch and Jacobs, 1979; Keller et al., 2000), motion discrimination(Douglas et al., 2006), orientation discrimination(Cowey and Franzini, 1979), and object recognition(Bussey et al., 2008; Minini and Jeffery, 2006; Zoccolan et al., 2009), but never on tasks with flanking stimuli. The presence of flankers made the task difficult for rats, presumably for both cognitive and perceptual reasons. In this study we are interested in the differential effects of the arrangement of flankers when they are present. The flankers' contrast, size and separation were held constant, while we varied their orientation, angular position and sign in randomly interleaved trials. We ask if rats' detection performance is sensitive to the relative orientation, position, and sign with respect to the target. We report an effect specific to collinear arrangements irrespective of sign.

## *Results*

**1. Rats can report the presence of a small oriented grating in the presence of flankers**

We developed an automated method to train rats to perform two-alternative forced-choice (2AFC) visual tasks (see Methods). We trained 7 male Long-Evans rats to detect an oriented grating target. The target was presented in the middle of a CRT display, and subjects were required to select one of two response ports to indicate that the target was either present or absent (Fig. 1a,b; photograph in Supplementary Fig. S1).

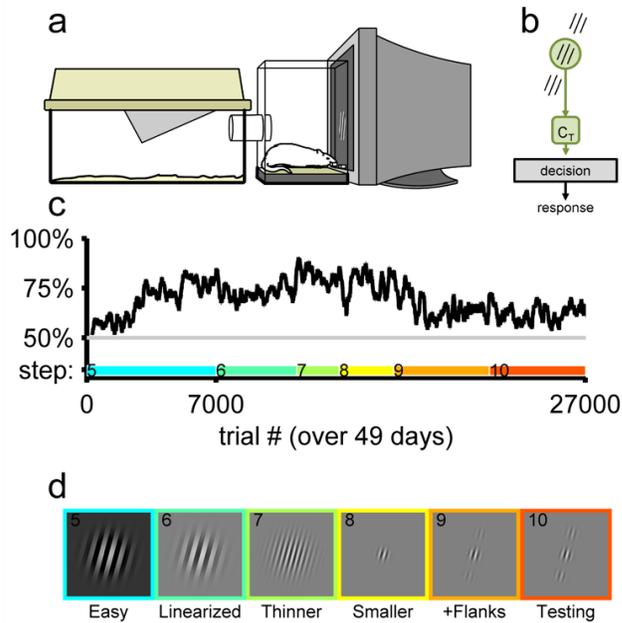

*Fig. 1. Training rats to detect an oriented target in the presence of flanking distractors.* a) A diagram of the training environment. Rats could initiate trials by licking a sensor centered in front of the display monitor. This immediately rendered a stimulus. Rats were rewarded with water for correctly reporting the presence or absence of a target by licking one of two response spigots located on the left or right side of the chamber. b) A simple schematic indicating the final task the rat is being shaped to perform. The target location is denoted by a green circle. The target is present on 50% of the trials; the target contrast, $C_T$, is either 0 or 1. The target's presence, irrespective of the flanker configuration, informs the rat which decision will result in a water reward. c) To achieve the final task which has small, low contrast targets and distracting flankers, rats are shaped through a sequence of training steps that increase in difficulty. Performance for a single rat is plotted as a 200 trial running average, starting from the first easy visual trials (step 5) to the testing phase (step 10). The first four steps involved associating the response ports with rewards, and did not involve any visual stimuli (see Supplementary Table S1 for details). When the rat's performance exceeded a preset criterion (>85% correct), he was automatically graduated to a new training step to ensure rapid progression and avoid over learning. d) Sample stimuli from each step are color-coded to match the performance plot, and named to emphasize the change from the previous training step. The addition of dim flankers (step 9) is displayed for a linearized contrast of 20% in b) but was increased from 10% up to 30% in step 9. All testing was performed with 40% linearized flanker contrast. For simplicity only one of the two orientations is shown, but all rats had equal training exposure to both. For more testing stimuli, see Figure Fig. 3. For a photograph of a rat performing the final task, see Figure S1.

The orientation of the target was tilted either clockwise or counter clockwise from vertical by a fixed angle; orientation was randomly chosen on each trial. Rats advanced automatically through a series of training steps that decreased target contrast, reduced its size, and increased its spatial frequency (Fig. 1c and d, steps 5-8). After rats learned the basic detection task in the absence of flankers, a brief testing period assessed the influence of the target's contrast and spatial frequency was tested on a subset of the trained rats (Supplementary Fig. S4a,c). This identified suitable parameters (see Supplementary Data 1) for the subsequent more difficult task involving distracting flankers.

Next we added two "flanking" gratings on either side of the target location (Fig. 1b, Fig. 1d steps 9-10). The target was absent on 50% of trials. Flankers were absent on 5% of the trials. During testing (step 10), flanker contrast was fixed at 40% and spatial parameters of the stimuli were independently varied ($\theta_T$, $\theta_F$, $\omega$, $S_T$, and $S_F$, as described in Fig. 3). Rats learned to perform target detection even in the presence of distracting flankers. We collected >20,000 trials per rat over 2-5 months of the testing step, which are further analyzed and summarized in Figures 4 and 5, and Supplementary Figures S3, S5-S7. Throughout the testing step, performance on trials with flankers remained well above chance (e.g., step 10 in Fig. 1c). Performance was stable over the period of data collection used for analysis (see Experimental Procedures).

The presence of flankers made the task substantially more difficult for rats (Supplementary Fig. S3). This effect is significant for 7 of 7 rats *individually (Agresti-Caffo 95% confidence interval) and* the population as a whole ($p<0.01$ on 2-way ANOVA; $p<0.01$ on Friedman's). Presumably this is due to both cognitive and perceptual factors, which we have not disambiguated here, but are considered elsewhere(Meier P, 2010). A perceptual effect on detection could arise from spatial contrast normalization, a form of surround processing long-observed in other mammals (see Introduction). Flankers add contrast to the target's surround, which could lower the target's effective contrast through contrast normalization. We verified that lower target contrast impairs detection as expected (Supplementary Fig. S4c), so detection should be sensitive to reduction in effective contrast. Flankers of higher contrast or closer proximity to the target should exert stronger contrast normalization, further reduce effective target contrast, and lead to larger impairments. Additional tests on a subset of rats confirmed both predictions (Supplementary Fig. S4b,d). Thus our task is a promising candidate for revealing effects that might depend on contrast normalization.

As indicated in the schematic of Fig. 1b, the rat's decision and response in our task obviously depends on contrast at the target location. A more complete schematic (Fig. 2) indicates that rats' decisions are also influenced by the presence of flankers. A contrast normalization component is indicated on the basis of past literature and in consistency with the data summarized above. The presence of flankers also affects performance for uncharacterized cognitive reasons, such as task confusion and compensating strategies. If performance is insensitive to the position and orientation of the flankers, a simple model like this would be sufficient.

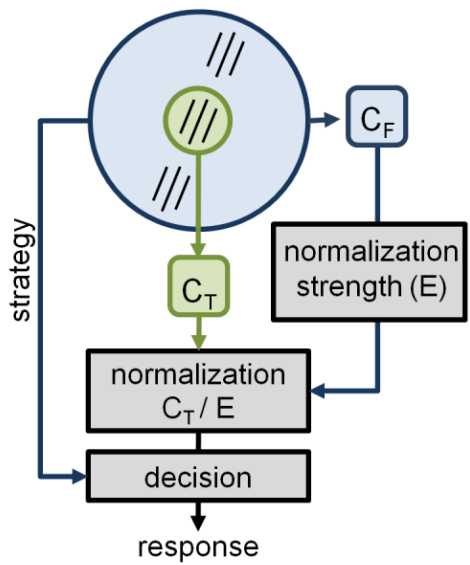

*Fig. 2.* *A schematic model indicating how the presence of flankers might influence rat's decisions. On this model, the contrast of the flankers ($C_F$) contributes to the normalization strength (E) such that the effective contrast of the target is reduced. The presence of spatial contrast normalization probably contributes to the deficits in performance associated with the presence of flankers. The presence of flankers also influences performance for cognitive reasons ("strategy"). This model is blind to orientation and position of flankers.*

## 2. Collinear flankers impair detection more than other arrangements

We next considered how performance depended on the arrangement of the flankers with respect to the target. There were two possible target orientations, as during training. Flankers always shared the same orientation and sign as one another, and were located symmetrically on either side of the target location on an imaginary line tilted either clockwise or counter clockwise (Fig. 3a). The target orientation ($\pm\theta_T$), flanker orientation ($\pm\theta_F$), and angular position of the flankers ($\pm\omega$) were chosen independently each trial for a total of 8 randomly interleaved stimulus configurations (Fig. 3b). The luminance signs of both target and flanker gratings ($\pm S_T$, $\pm S_F$) were also randomized for each trial.

We use the term "collinear" to refer to stimulus configurations in which the target and flanker orientations are both aligned with the flanker angular position ($\theta_T = \theta_F = \omega$), irrespective of the relative sign. This configuration could engage visual processing that relates line segments that fall along a common contour. We label non-collinear conditions as follows: "popout$_1$" ($\theta_T = -\theta_F = \omega$), "popout$_2$" ($-\theta_T = \theta_F = \omega$) and "parallel" ($\theta_T = \theta_F = -\omega$). For examples, see Figure 3.

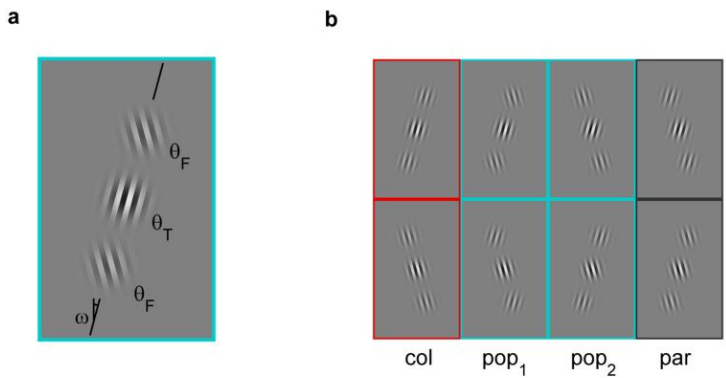

*Fig. 3. Grouping flanker stimuli into conditions.* a) An example flanker stimulus, with labeled spatial parameters ($\theta_T$, $\theta_F$, $\omega$). Flanker stimuli where generated by independently varying target orientation ($\theta_T$), flanker orientation ($\theta_F$), and the angular position of the flankers ($\omega$). The two flankers always had the same orientation. b) The different stimuli were grouped in four conditions that preserved geometric relationships. The top and bottom row are mirror images of one another. In the collinear condition, the target and flanker orientation align with the angular position ($\theta_T = \theta_F = \omega$). Collinearity was disrupted by changing one of the parameters in each of the remaining flanker conditions: "popout$_1$" ($\theta_T = -\theta_F = \omega$), "popout$_2$" ($-\theta_T = \theta_F = \omega$) and "parallel" ($\theta_T = \theta_F = -\omega$). Each condition was equally likely during the testing. Every stimulus has a matching case in which the target was absent (not shown here, but see Supplementary Fig. S2b). During all training and testing, the luminance sign of the target and flankers were randomized; the case of $S_T=S_F=+1$ is shown.

The main finding of our study is that the collinear condition is consistently harder for rats than any of the other three configurations (Fig. 4a,b). This difference was true for each rat and significant at the population level even when adjusting for multiple comparisons (p<0.01 by Tukey-Kramer on 2-way ANOVA for all three comparisons; two of these comparisons were also significant by the more conservative Tukey-Kramer on Friedman p<0.01; see Supplementary Data 3, Supplementary Fig. S6). The other three conditions were not significantly different from one another (Supplementary Fig. S5,S6). In short, of all the flankers conditions tested, only the collinear condition was consistently most difficult.

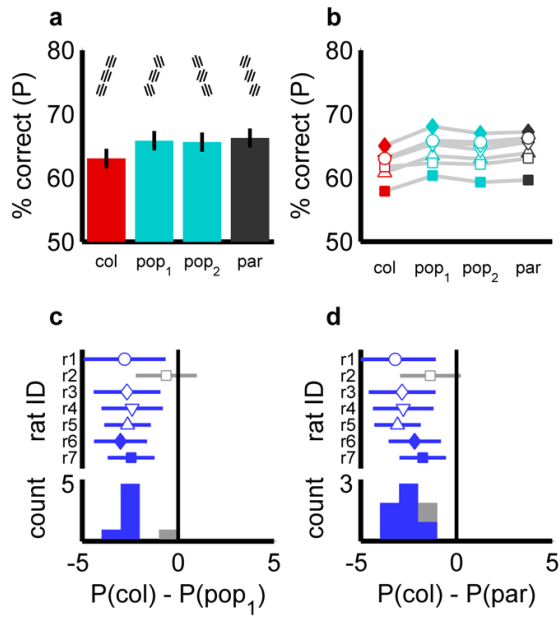

*Fig 4. Collinear flankers impair detection more than other arrangements. a) A single rat's performance (r1) on the four conditions: collinear, and three patterns that disrupted collinearity. In **pop-out$_1$** collinearity was disrupted by changing the flanker orientation ($\theta_F$) to be different from the target($\theta_T$). **Pop-out$_2$** maintained the same difference in flanker and target orientations ($|\theta_F-\theta_T|$) as popout$_1$, but the angular position of the flanker was different ($\omega$). In **parallel** the collinearity was disrupted only by changing the angular position of the flanker. Error bars indicate 95% binomial confidence intervals. b) Performance of all seven rats on all four conditions. c) The difference in percent correct between collinear and pop-out$_1$ for all rats (r1-r7). Collinear is more difficult. Error bars indicate 95% confidence interval using a modified Wald interval described in methods (Agresti and Caffo, 2000).One rat's data are rendered gray to indicate that the difference in performance is not significant. The other six rats are rendered blue because they are each significant(Agresti-Caffo 95% confidence interval). Filled symbols indicate subjects in which the mapping between yes-no and left-right was inverted (See training protocol in methods). Subjects with inverted training rules had no different effects. d) The difference in percent correct between collinear and parallel. Again collinear is harder. For both comparisons (panels c and d), the difference between conditions is significant for six of seven rats individually (Agresti-Caffo 95% confidence interval), and for the population as a whole(2-way ANOVA with Tukey's, p<0.01). . For all possible pairwise comparisons, see Figure S5.*

All stimulus arrangements had the same flanker contrast and distance, so this difference cannot be explained by simple contrast normalization as illustrated in Fig 2. We consider next how the contrast normalization framework could be extended to account for a collinear deficit at the level of early visual proecessing. The collinear flanker condition differs from the popout$_1$ condition only by flanker orientation, so the difference in performance (Fig. 4c) suggests orientation-sensitive surround processing. This could be explained by a simple modification of the model in Figure 2, such that the strength of contrast normalization (E) is stronger when the contrast in the surround shares the target's orientation. Such a model could account for our result that the collinear flanker condition was harder than either popout condition, but could not explain why performance in the parallel condition was significantly better than collinear (Fig. 4d), and indistinguishable from either popout (Supplementary Fig. S5). Our data cannot be explained by simple orientation-dependent effect, nor by an angular position effect alone. Feature arrangement is important: flanker orientation and angular position interact. To capture differences between flanker conditions within the perceptual component of the model, it would be necessary to add a pattern-sensitive term (see Discussion, Fig. 6).

Flankers that are collinear to the target either had the same luminance sign ($S_F = S_T$), such that white bars of the target align with white bars of the flankers, or opposite signs, such that white bars align with black bars ($-S_F = S_T$). A reversal in sign is equivalent to a $\pi$ shift in spatial phase. We could find no effect on performance of the relative or absolute luminance sign of the flanker and target gratings. In particular, whether the dark bars in the target aligned with the dark or light bars in the flankers, the specific impairment for the collinear configuration relative to other arrangements remained, and the amplitude of the effect was indistinguishable ($p > 0.05$, 2-way ANOVA; $p > 0.05$, Friedman's test; Supplementary Fig. S7). In a pilot study, intermediate phase shifts also had no effect in a related task (Supplementary Fig. S7). Therefore we do not include phase as a parameter in Figure 6.

The arrangement of nearby features has been shown to affect behavior and early visual processing in other species (see Introduction). This is the first demonstration of pattern sensitivity in a rodent, showing that such effects can occur even in species that lack orientation columns. This is also the first flanker study in any species in which both orientation and position were randomized in a single interleaved testing period, confirming that neither a position effect nor an orientation effect are sufficient to explain the collinearity effect.

## 3. Collinear flankers cause rats to miss target

Performance reflects both the ability to say yes when the target is present (hits) and the ability to say no when the target is absent (correct rejections). We find the collinear condition decreases accuracy of both kinds. For each rat, the hit rate was lower for the collinear condition than the parallel condition (Fig. 5a; vertical axis in Fig. 5c,d). The false alarm rate was also higher for the collinear than the parallel condition (Fig. 5b; red horizontal axis in Fig. 5c,d). The average decrease in hit rate (3.7%, Fig. 5a) was about three times larger than the average increase in false alarms (1.1%, Fig. 5b). The decrease in hit rate is significant for 6 of 7 rats (Fig. 5a) whereas the increase in false alarms is significant for only 1 of 7 rats (Fig. 5b). The same trends are found when comparing collinear flankers to either popout condition (not shown). The net effect is that collinear flankers cause rats to report "no" more often than other flankers. They cause rats to miss the target.

        The hit rates and false alarms for collinear and parallel conditions are also shown in an ROC space (Fig. 5c). Data from the one outlying rat (r2) is included next to a rat displaying a typical effect (r5) in the expanded view (Fig. 5d). Although subjects differ in overall performance and bias, the effect of collinearity is similar for all subjects, indicated by the consistent direction of the arrows. The increase in misses is the dominant effect in the population.

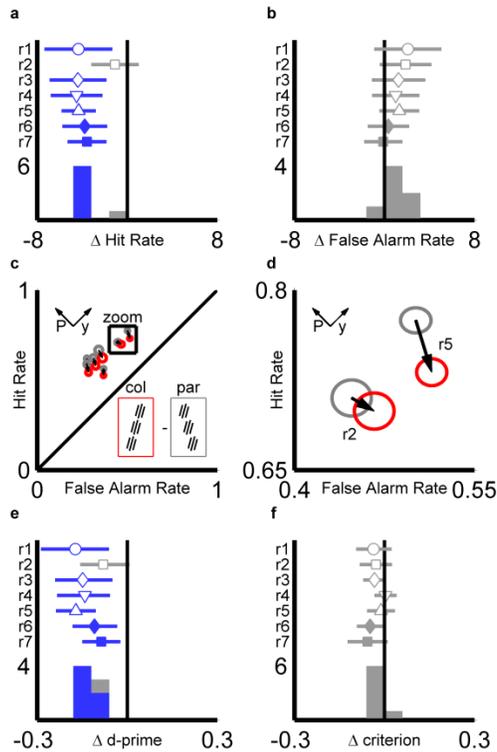

*Fig. 5. **Collinear flankers cause rats to miss target**. A) The change in hit rate between collinear and parallel configurations, h(col)-h(par). Symbols left of the zero line indicate that the hit rate is lower when flankers are collinear. The effect is shown for each subject: the symbol location indicates mean difference in hit rate, horizontal lines indicate 95% confidence interval(Agresti and Caffo, 2000). For all but one rat, the deficit with flankers is statistically significant. b) The change in false alarm rate, f(col)-f(par), using the same conventions as panel a. Rats display more false alarms on the collinear stimuli. However, the change in false alarms was smaller than the change in hit rate show in panel a. c) A geometric representation of the raw data in a and b is shown by plotting the Receiver Operator Characteristics (ROC). The individual ellipses show the boundary of the 95% confidence intervals of a binomial distribution for hit rate and false alarm rate for one subject and condition. For each subject, an arrow in ROC space summarizes the difference between the collinear condition (red) and randomly interleaved trials of the non-collinear reference condition (parallel, gray). The arrow points to the change in responses induced by the collinear feature with respect to the reference condition. Most arrows consistently point down and to the right indicating a decrease in hits, and a small increase in false alarms. d) Detail of subpanel c to provide better resolution for a typical rat (r5) and an outlier rat (r2). Interestingly, histology (not shown) revealed that the rat displaying the outlier effect (r2) had a naturally occurring tumor which compressed and displaced about a quarter of the ventral thalamus. We do not know if this played a role in the animal's behavioral differences. e) The difference in detection sensitivity between collinear and parallel conditions. Errorbars are the 95% confidence intervals of samples drawn from a Monte Carlo Markov Chain. These d' measurements are consistent with the report in Fig. 4d: collinear stimuli have targets that are harder to detect. f) The change in criterion between collinear and parallel conditions. Negative values indicate that rats are more likely to report the absence of the target in the collinear condition. The change in bias is consistent for all rats, but not significant for any individual rat (MCMC 95% confidence interval). The size of the bias difference is small compared to the change in d' shown in panel e.*

Signal detection theory interprets these raw data in terms of sensitivity (d') and bias (criterion). Applying this framework, sensitivity is consistently lower when flankers are collinear (Fig. 5e). This effect is significant for 6 of 7 rats (Agresti-Caffo 95% confidence interval) and for the population as a whole ($p<0.01$ by Tukey-Kramer on 2-way ANOVA). Rats also show a consistent shift in criterion, reflecting the greater bias to say "no" for collinear stimuli (Fig. 5f). The criterion shift is small compared to the change in sensitivity, and is not significant for any rat. We cannot confirm the assumptions of Signal Detection Theory hold in our study, but our conclusion (that collinear flankers cause misses) is observable in the raw data (Fig 5a-d) independent of these assumptions.

## Discussion

These data show that detection of visual stimuli by rats is sensitive to the configuration of the flanking elements. In particular, flankers collinear to the target impair performance compared with other configurations. Agreement in sign between target and flanker gratings was not required for this effect. This result suggests specialized processing of oriented image features that can be connected to form a continuous contour. It is noteworthy that this processing must occur in the absence of orientation columns, which are absent in rats(Ohki et al., 2005).

Contrast normalization is a powerful conceptual framework for explaining many surround effects in early visual processing. A pattern-sensitive generalization of contrast normalization could account for a collinear effect (Fig. 6). In this model the normalization strength (E) includes additional dependencies on the parameters of spatial configuration ($\theta_T, \theta_F, \omega$). This extension of the model in Figure 2 allows the normalization strength (E) to be specific to orientation in the target location, and sensitive to the specific arrangement of flanking features. The dominant effect of pattern in our data could be explained by a single factor that selects for the alignment of all three experimental parameters: $\theta_T = \theta_F = \omega$. In this model, collinearity increases the normalization strength leading to greater performance impairment. Other perceptual and/or cognitive models could also account for our behavioral data, if they incorporate a "collinearity" term sensitive to the interaction of position and orientation of flankers. Collinear effects could be ascribed to higher visual processing areas or cognitive effects. For example mechanisms for binding features, processing gaps, or interpreting occlusions could play a role. Here we offer one plausible and parsimonious model which makes direct predictions that can be tested physiologically.

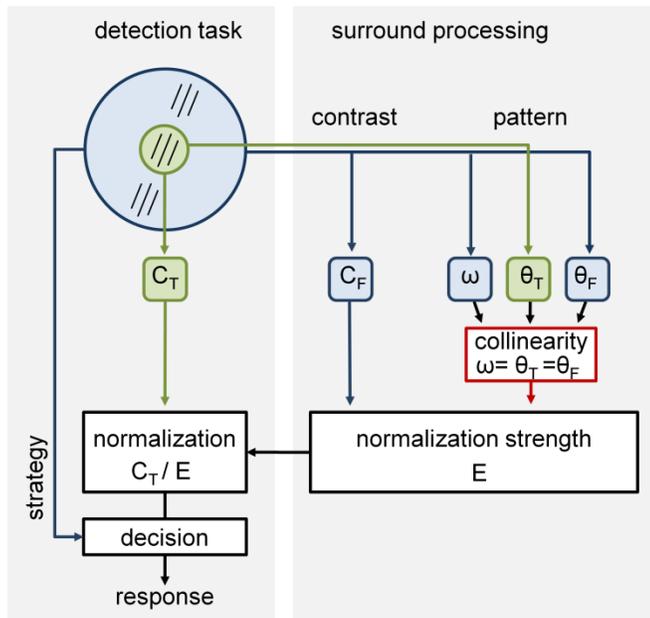

*Fig. 6. Schematic model of pattern sensitive contrast normalization. The detection task is summarized in the left hand region, unchanged from Figure 2. The contrast in the target region ($C_T$) is represented by a neural signal which is normalized by surround processing in early vision before the decision is made. Then the rat responds left or right to indicate the target's presence or absence. In this model, cognitive effects of the surround ("strategy") depend only on the presence of flankers and insensitive to their configuration. Surround processing contains two aspects: sensitivity to contrast, and sensitivity to pattern. In this study, surround contrast is determined by the experimental parameter for flanker contrast ($C_F$, same as Fig. 2), and was held constant during testing. The pattern-sensitive component must include at least three terms to account for our data: the angular position of the flanker ($\omega$) the orientation of the target ($\theta_T$), and the orientation of the flanker($\theta_F$). While there are many ways that these terms could interact, we only require a dependence of collinearity to explain the rat's behavior. The processing of contrast and pattern in the surround is combined into a single normalization term (E). This determines the gain of the neural signal that is used for detection. An argument that the normalization term E could be interpreted as an expected contrast ($\hat{C}_T$) is considered in the discussion.*

Collinear flankers cause rats to miss targets more than other flanker configurations (Fig. 5a). This is consistent with perceptual masking (rats not seeing the target), but we are reluctant to attribute the bias change to a perceptual process alone or a decision process alone without a measurement of an internal signal that represents the target. For example, if rats suppress perception of a target due to a lateral mask, or perceive a false target due to an illusory contour, they might learn to shift a downstream decision criterion to a new boundary that maximizes their reward rather than reporting their percept. Therefore we do not think the change in bias observed in our data supports any strong conclusions about perception. Nevertheless the data may constrain future models, so we report the raw values we observed for all conditions and rats in Supplementary Analysis 1 (Supplementary Fig. S8 a,d,g).

**Potential Confounds**

Our data show that rats' behavior is sensitive to the arrangement of oriented visual features above and beyond the effects of nearby contrast. In any flanker psychophysics study, one should address confounds that might arise from slow variation in performance, familiarity with stimuli, response biases, stimulus artifacts, cognitive confusion, or the influence of attention. Here we address each of these potential confounds.

During the test period, each rat's performance was approximately stable. Of course, performance does fluctuate, probably due to slow variations in motivation. Also, we cannot exclude a small effect of expertise learning. These correlations over time could influence blocked performance such that temporal variation would appear as differences across experimental conditions. For this reason we randomly interleaved all condition types using the method of constant stimuli. This method also balances subjects' exposure to long runs of the same flanker configuration (Fig. 3b), which may be difficult to accomplish with adaptive psychometric methods.

Second, it is possible that the exposure to certain orientations, in the recent past, or throughout a subject's lifespan, could influence their perceptual processing of that orientation(Kurki et al., 2006). If we had only used a single target orientation in training or testing, we could not rule out effects of orientation-specific familiarity or adaptation. Therefore subjects were exposed to the same distribution of target orientations and signs in the training steps as in the testing step. Moreover, whenever flankers were present they had the same distribution of properties as the target.

Third, if the rats' 'yes' and 'no' behavioral responses are inherently asymmetric, this would complicate interpretation. We avoided a go-no-go trial structure because it is likely different circuitry is required to initiate versus inhibit a response. Instead we used a 2-alternative forced-choice trial structure where both 'yes' and 'no' required initiation of a symmetric motor output. Reinforcement was also symmetric: correct trials were always rewarded and incorrect ones always initiated a time out, regardless of the target's presence. As a further control, two subjects (r6 and r7) used the same experimental equipment as their brothers, but were trained with an inverted rule, such that "yes" and "no" were mapped onto the opposite sides. The results for these rats were not different (Fig. 4 and 5).

Fourth, target orientation or flanker properties might affect target visibility through artifacts of the monitor, rather than processing in the brain. Specifically, is it known that vertical gratings presented on any CRT monitor have lower effective contrast than horizontal stimuli, because RGB guns follow rasterized horizontal scans and lack perfect temporal resolution(Garcia-Perez and Peli, 2001). Had we used gratings that weren't symmetrically tilted about vertical, these artifactual differences in contrast could have been responsible for performance differences across conditions. We also designed the stimuli so that flankers and targets never shared horizontal scan lines to minimize their impact on each other's contrast.

Fifth, it is possible subjects' errors are not due to perceptual difficulty, but rather a failure to understand the intended task. We confirmed in 2/2 rats that their detection performance in all flanker conditions was sensitive to the target's contrast (Supplementary Fig. S4c). Because their performance did not saturate with the contrast we used in our study, at least some incorrect responses were due to perceptual difficulty. We cannot rule out that cognitive difficulty may also have contributed to errors. For example, the decrease in performance when flankers are added (Supplementary Fig. S3) could be explained if rats failed to understand that the target location contained the relevant feature, and also responded to gratings in non-target locations. However, this confusion would not explain the consistent collinearity impairment observed in all individuals (Fig. 4 ).

Finally, spatial or feature-specific attention may play a role in some flanker tasks(Freeman et al., 2004). In our task, to ensure that feature-specific attention would not give the rats a differential advantage between stimulus conditions, target orientation was randomly chosen each trial. We did not employ a positional cue for target location because forward masking could affect target detection. In our task, flankers could improve allocation of spatial attention by reducing uncertainty about the target's location, ultimately improving performance compared to trials without flankers(Petrov et al., 2006). This could occur in our task, but if rats did benefit from spatial uncertainty reduction, other effects of the flankers overwhelmed this benefit, yielding net decreases in performance. Alternatively, rats' attentional allocation or visual representation might lack the spatial resolution to isolate flankers from the target location.

These factors could explain the flanker-induced impairment (Fig S3), but not the collinear specificity (Fig. 4).

In summary, we have controlled for slow variations in behavior, balanced stimulus familiarity, used symmetric responses, avoided CRT artifacts, confirmed that targets are perceptually challenging to detect, and avoided confounds due to orientation-specific attention. We conclude that the rat visual system is sensitive to pattern above and beyond the effects of nearby contrast. We attribute this sensitivity to the rats' visual system, as opposed to other sources of variability in the environment or the rats' cognition.

**Comparison to humans**

In some perceptual tasks, the presence of collinear flankers improves human performance(Chen and Tyler, 2008; Polat and Sagi, 2007). Why did the collinear flankers impair behavior in rats as opposed to improve it? The term "facilitation" and "suppression" refer to either increases or decreases in performance at a fixed contrast, as in this study, or the ability to match a constant performance in a new condition using a lower or higher target contrast. In this study collinear flankers suppressed detection in rats. On the other hand, collinear flankers facilitate detection for humans performing a two interval forced choice task in the lateral masking paradigm(Polat and Sagi, 1993; Solomon and Morgan, 2000; Williams and Hess, 1998) and dual masking paradigm(Chen and Tyler, 2008). We note that in different task paradigms, this facilitation is not found. The human study most similar to ours also used randomly interleaved trial conditions, fixed contrasts, a single stimulus yes/no paradigm, and oblique vs. collinear flankers, but found no collinearity effect at the target-flanker proximity we used (their Fig. 6, 3λ)(Polat and Sagi, 2007). It remains to be determined whether this difference is attributable to a difference in species, training experience, or stimulus parameters: they used sinusoidal gratings and a single target orientation.

Human studies show that flanker effects change in magnitude or even sign, depending on the contrast regime - which is low contrast for detection tasks and higher for contrast discrimination. The pattern of results in both can be cast in a contrast normalization framework, though the different contrast regimes may involve disparate cellular and circuit mechanisms. The results above were obtained using a target contrast of 1.0 (where contrast is reported as the fraction of the linearized range of the display spanning 4-42cd/m$^2$). This contrast is near detection threshold for rats at the spatial frequency we used (Supplementary Fig. S4a). Collinear facilitation is reported to be strongest at lower target contrasts. Thus we also analyzed the data collected from three lower contrasts (25%, 50%, 75%). In no case did any target contrast or any configuration of flankers improve detection compared to the target alone condition (data of Fig. S4c, analysis not shown).

In our data we did not observe sensitivity to the relative sign of target and flankers. The two signs we used are equivalent to having one of two spatial phases. In human psychophysics, both phase-sensitive(Chen and Tyler, 1999; Ejima and Takahashi, 1985; Williams and Hess, 1998; Zenger-Landolt and Koch, 2001; Zenger and Sagi, 1996) and phase-insensitive(Chen and Tyler, 1999; Field et al., 1993; Xing and Heeger, 2001; Zenger and Sagi, 1996) collinearity effects have been described, perhaps reflecting the phase-sensitive (simple cell) and phase-invariant (complex cell) processing channels for orientation in V1. Differences among these studies that appear to be relevant include the distance of flankers from target (gap, no gap, or overlap), and whether the stimuli are presented in the fovea or periphery(Chen and Tyler, 1999). In rats, we only tested one distance with no overlap ($3\lambda$), one spatial frequency (0.22 cyc/deg), and two phases (aligned and reversed). We do not know what part of the retina rats used in the task, nor whether rats' central vision(Heffner and Heffner, 1992) is more similar to foveal or peripheral vision in primates. While it may seem that phase sensitivity is useful for pattern processing, human psychophysics suggest invariance to phase is a hallmark feature of contour integration(Williams and Hess, 1998).

Though flanker effects reported in the human literature depend on stimulus and task details, our results agree with the consistent key finding across human studies: performance under collinear conditions is special.

**Relating our findings to natural scene statistics**

There is a substantial literature theorizing that early visual processing is optimized for the statistics of natural scenes(Barlow, 2001; Field, 1987; Geisler, 2008). These optimizations can impair performance in tasks that violate natural scene statistics(Howe and Purves, 2005; Schwartz et al., 2009; Weiss et al., 2002). In light of this theory, it is noteworthy that the condition that most affects rats' target detection, collinear flankers, corresponds to the feature conjunction that is statistically most frequent in natural scenes. Combining this perspective with a contrast normalization model leads to a speculation about how a pattern-specific normalization pool could be acquired by learning, without requiring anatomic segregation of orientation channels.

Suppose the visual system computes a prediction of target probability based on some function of the image at other locations, and this prediction is used to adjust the local representation of target. The theory implies that those surrounds that make the strongest predictions about the target in natural images should influence the representation most, and thus impair performance most in our task. This theory is neutral about whether predicted targets should be suppressed (reducing redundancy) or enhanced (propagating beliefs) at the level of early vision. The direction of this influence cannot be predicted and may be species specific. The framework of normalization developed above implies suppression of predicted features.

It is biologically plausible that the visual system could perform this computation. For example, sensitivity to the separate pair-wise correlations of nearby local oriented features might be learned from the correlated firing of V1 neurons by activity-dependent mechanisms, without requiring orientation columns. Suppose that each local oriented feature's representation is normalized by the activity of all nearby local oriented features, in proportion to their statistical co-activation in natural images. In the statistics of natural images, contrast at one location is correlated with high contrast nearby(Ruderman and Bialek, 1994). The co-occurrence of oriented features depends on the relative orientation and position, and collinear features co-occur most often(Geisler et al., 2001; Itti and Koch, 2000; Sigman et al., 2001). Thus all flankers should normalize, and collinear features should normalize the most(Schwartz and Simoncelli, 2001). In our task, we find that all flankers impair detection and collinear flankers impair the most.

In this framework, one can think of the normalization strength in our model ($E$) as representing a predicted contrast at the target location ($\hat{C}_T$) estimated on the basis of the contrast in the flanking region ($C_F$). This suggests that the function of divisive normalization is to reduce the effective contrast for expected features, and amplify unexpected features, thereby maximizing information transfer for natural scenes(Ruderman and Bialek, 1994).

Future studies could further test this ecological interpretation by correlating behavioral impairment with natural co-occurrence statistics of flanking features at other positions and orientations. In particular, it will be of interest to explore parallel flankers at other positions, and popout flankers with greater orientation differences. The more specific hypothesis of normalization makes the direct prediction that flanker features that are correlated with the target in natural images should reduce the firing rate of neurons that respond to the target early in the visual system (such as thalamus or primary visual cortex). Surround processing has not been studied in these neurons in rodents. In cat and primate V1, surround stimuli generally reduce firing(Bonds, 1989; Carandini et al., 1997; Cavanaugh et al., 2002; Heeger, 1992; Polat et al., 1998; Shapley and Victor, 1979), but some cells fire more when flanked by collinear features(Li et al., 2006; Polat et al., 1998; Sillito et al., 1993).

Contrast normalization would reduce redundancy, leading to more efficient codes, but this is not the only goal of vision. Pure contrast normalization may even be at odds with other visual goals. Statistical inference from surround stimuli could contextually de-noise the signal, leading to better parameter estimation through the combination of weak signals. This would also exploit the correlated signals of the natural world(Barlow, 2001) but with opposing effects. We presume that both processes occur and interact in natural vision; different tasks may emphasize one or the other. We focus on the role of contrast normalization because it requires the least complexity to explain all of our data.

**Rats as a model system for vision research**

The impact of flankers on behavior has only been studied in humans and other primates, the physiology of the early visual system primarily in non-human primates and cats. Rats offer several advantages as a vision model. Rat husbandry is inexpensive and their behavior, neuroanatomy, and neurophysiology are extensively studied. We have demonstrated that they are easily trained to perform visual tasks that involve distracters and that their vision is sensitive to the spatial arrangement of features. This adds to the growing list of visual tasks demonstrated in rats(Birch and Jacobs, 1979; Cowey and Franzini, 1979; Douglas et al., 2006; Keller et al., 2000; Zoccolan et al., 2009). In this study, individual rats performed around 500 trials every day with stable performance over months and require little human supervision. Many powerful techniques are more feasible in rats than primates or cats, such as genetic, transgenic, viral, histological, optical, intracellular, and pharmacological methods. We conclude that rats provide a valuable complementary model system for studying contextual visual processing.

# Experimental Procedures

**Animal subjects**

Data were collected from seven male Long Evans rats (Harlan Laboratories). All experiments were conducted under the supervision and with the approval of the *Institutional Animal Care and Use Committee* at the University of California San Diego.

The rats included in this study were the 7 median performers from an initial cohort of 14 animals. Four animals were excluded from this study because they either remained at chance on the initial learning task or their performance never exceeded our automatic graduation criterion. They never saw flankers. Three of the remaining ten rats were high performers and were moved to another study before collecting a sufficient amount of data on the testing step. They performed 2, 7 and 16 sessions while other rats performed 60-150 sessions.

**Training System**

We designed custom hardware and software for automation and parallelization of training. A broad overview of its design and architecture can be found in Supplemental Experimental Procedures 2. Each station consists of a CRT display adjacent to a transparent cage that interfaces easily with slightly modified standard vivarium rat cages (Fig. 1a). Rewards were spatiotemporally co-localized with response. Our initial training methods were adapted from previous studies of olfactory and auditory tasks in rats(Otazu et al., 2009; Uchida and Mainen, 2003). Also see other similar methods. (Zoccolan et al., 2009). In our study. behavior was detected via three infrared beam break detectors (Optek OPB980T11) in stainless steel housings for protection against chew damage. Water was delivered to a rounded 16 gauge stainless steel tube positioned just behind each beam by computer-timed solenoid valve opening (80 ms, Neptune Research, 161PO21-11, 161T01, Cooldrive) from a pressurized source (~300 mmHg, Infu-surg 4010, Ethox Corp.) through rigid tubing ($CO_2$ lines 8044, SurgiVet). Our reward volume was roughly 16 μL, and was typically delivered within 10 ms of a correct lick response. Occasionally rewards were larger, described in training protocol. Auditory feedback was provided with ear-bud headphones mounted on the side walls of the box. Different sounds indicated when a detector beam was broken, and differentiated responses as request, correct, incorrect, and inappropriate (left- or right- licks with no preceding center request lick). To present visual stimuli, we used PsychToolbox(Kleiner, 2007) to control standard OpenGL capable graphics cards (Nvidia GeForce 7600) via Matlab (Mathworks, Natick, MA). We did not track head position or gaze. The head position, head orientation, and eye level and is fairly consistent from trial to trial within subject, as determined from direct observation.

**Training Protocol**

We designed a series of shaping steps that gradually taught rats to perform the detection task with flankers. An overview of the steps is provided in Supplementary Table S1. Details of the general training procedure such as water restriction, schedule, and environment can be found in Supplemental Methods 1. The specific shaping sequence used for the subjects in this study was as follows:

*Learning to lick.* The goal of the first three steps was to teach the rats to use the detector/reward ports. We presented no visual stimuli, and rats obtain a reward by licking any port at any time. To encourage rats to move among the ports, only one consecutive reward was allowed per port. During step 1, the system occasionally stochastically generated a drop of water without any action from the rat, in order to generate interest in the ports. On step 2, these automatic rewards were turned off, so that rats must actively lick the ports. Rats graduated by sustaining 5 rewards per minute for 2 minutes. Step 3 was identical to step 2 except it required a stricter graduation; it need not be included in future studies. All rats in this study passed through step 3 in 1-2 sessions, with the exception of one rat that got stuck on step 3 for 6 sessions.

*Learning two-alternative forced choice (2AFC) task structure.* From step 4 onward, trials had a 2AFC structure. A center lick initiated a trial (but was not rewarded), and the first subsequent lick at either the right ("target present") or left ("target absent") port determined the trial outcome. For 2 of the 7 rats, the present/absent port identities were reversed. Targets appeared on 50% of trials; for these trials, a response at the "present" port was rewarded. Otherwise a response at the "absent" port was rewarded. The tone that accompanied each trial request provided the subject with confirmation that he had successfully initiated a trial and should proceed to ascertain and report target presence; a different tone and a flickering screen accompanied errors for the duration of the time-out penalty indicating the system was non-responsive. The gratings appeared on the same grey field background that was displayed between trials (after a response, during a reward, and before the next request).

We used two techniques to discourage guessing, side-biases, and other undesirable behavior patterns. First, 50% of incorrect responses were followed by "correction trials." During correction trials, trials with new stimuli but the same correct answer are repeated until the rat responds correctly. Correction trials can induce a rational strategy of switching response after an error (if 50% of errors are followed by correction trials, always switching responses after an error gives 75% performance for those trials – above chance performance without reference to the stimulus). The rats did, in fact, bias their responses in this way, so trials immediately following errors, including correction trials, were omitted from analysis. Strategies that ignore the visual stimulus can only impair performance in any trial that did not follow an error trial.

Second, we gave increasing rewards for consecutive correct answers. The first correct response after an error yielded an 80 ms valve opening (approximately 16 µL). The $2^{nd}$ to $4^{th}$ consecutive correct responses earned 100, 150, and 250 ms rewards. Consecutive responses thereafter earned 250 ms rewards; the first incorrect response reset this schedule to the beginning value of 80 ms.

For step 4, the first step with visual stimuli, the target was a large full-contrast square-wave grating masked by a circular Gaussian (21.6° standard deviation from the reference viewing location, with spatial frequency 0.11 cyc/deg, Fig. 1d, same stimulus as step 5). We used a square-gratings instead of a sine-gratings because we speculate that contour integration mechanism might engage most strongly for sharp edges. The Gaussian was truncated at 4 standard deviations (about the limit of 8-bit discretization). The grating bars were slanted clockwise or counterclockwise from vertical with equal probability (±15° for r1-r5, ±22.5° for r6 and r7). Results from r1-r5 and r6-r7 were combined because the trends and effect magnitudes were not significantly different ($p>0.05$ t-test, $p>0.05$ Kruskalwallis). For this and subsequent steps, monitor resolution was set to 1024x768 pixels at 100 Hz.

Because this step was designed only to establish the 2AFC trial structure (center request followed by side response), the graduation criterion required only high sustained trial rates, *not* above-chance performance.

*Visual detection.* Step 5 introduced a penalty timeout period for incorrect responses (1-6 seconds, hand-tuned for each rat), during which the "incorrect response" sound played and rats could not initiate a trial. Graduation from steps 5-8 required 85% correct performance on the previous 200 trials or 80% on the previous 500 trials. Graduation from step 5 represents the first evidence that rats can perform visual detection in our apparatus (Fig. 1c).

Step 6 introduces a gamma correction table (previous steps leave the CRT's native gamma uncorrected) that linearizes the monitor's luminance output, so that the $2^8$ grey levels correspond to equally spaced increments in $cd/m^2$ (See Supplemental Methods 3). After linearization, stimuli used a smaller luminance range and had a higher mean luminance than the earlier training steps. This reduced the effective contrast of the stimulus. All subsequent steps used linearized stimuli.

On step 7, grating spatial frequency increased to 0.22 cyc/deg (equivalent to $\lambda = 4.5°$ where $\lambda$ is degrees per cycle, a standard unit for indicating flanker distance(Polat and Sagi, 1993)). Pilot studies (Supplementary Fig. S4c,d) identified this value as in the range where rats' detection performance was not saturated, but strongly sensitive to target contrast (ranging from 60-75% as target contrast ranges from 50%-100% of the linearized range). This was necessary in order to ensure that we could observe either performance improvements or impairments caused by flankers. This value is consistent with previous reports of rat acuity(Birch and Jacobs, 1979; Heffner and Heffner, 1992; Jacobs et al., 2001; Keller et al., 2000; Prusky et al., 2000), which starts to roll off at 0.1 cyc/deg and shows no sensitivity above about 1.0 cyc/deg for high contrast displays. Our Gaussian mask reduces contrast throughout the grating except at the centermost pixel.

Step 8 reduces the grating size by about 70%, so that the standard deviation of the Gaussian mask was 6.8°. Subjectively, this left about 3 visible periods in the grating. Flankers were not yet included, but at this grating size and spatial frequency, the monitor had room to display complete flankers (size and spatial frequency identical to the target) at a distance of up to $5\lambda$.

Flanking targets were introduced at a distance of $3\lambda$ during step 9. Their contrast was slowly increased from 10% to 40% with training; each 10% was added after a performance criterion of 80% was reached. Performance remained above chance for higher flanker contrasts, but these were not chosen for testing because it is difficult to resolve differences in performance between conditions less than 60% correct. The increasing impairment of performance with flanker contrast could reflect either suppression of the target's apparent contrast below a detectable level, or confusion of high contrast flankers with rewarded targets. For this flanker distance and Gaussian mask size, flankers were non-overlapping, and subjectively appeared separated from the target (Supplementary Fig. S1). For each trial, the target orientation ($\theta_T$), flanker orientation ($\theta_F$), and angular position ($\omega$) were randomly and independently chosen to be either ±15° (rats r1-r5) or -22.5° (rats r6 and r7).

*Testing Stimuli.* Test stimuli were the same as those in step 9 with 40% contrast flankers (Fig.3). For the purposes of analysis, we grouped all the flanker conditions into four categories depending on the relationship of the flankers to the target (Fig. 3b). During testing, error feedback and reward contingencies were not changed. Correction trials continued to occur after 50% of errors.

**Performance stability, data filtering, and performance measures**

Subjects performed 25,000-90,000 trials during the final testing phase, over the course of 67-136 training days. To assess the stability of performance over time, we calculated performance for each rat and condition in consecutive non-overlapping 500 trial windows. Considering the entire testing period, each subject's long-term performance trended slightly up or down, but this drift always amounted to <7% total change. The most unstable performance was exhibited by r5, whose behavior fell for unknown reasons from ~65% to ~55% for ~6000 trials (~12% of the data in his testing phase) and then recovered. Despite unstable performance over time, the influence of collinearity for this rat was typical of the population. The performance in 500-trial windows ranged between 52-72% for every rat and flanker condition, averaging 63% overall.

We excluded from analysis all trials following an error trial, because rats showed evidence of an alternation strategy after errors, perhaps due to correction trials (see above). Including non-correction trials that immediately followed errors reduces the effect of collinear flankers, but not below significance; this did not alter any trends or influence our interpretation. Considering only trials in the central 80% of reaction times can remove many aberrant trials where rats were either not on-task or very rapidly performing trials while apparently ignoring the visual stimulus. This improves performance and makes the collinearity effect appear stronger, but we do not filter the data in this way for the purposes of this publication.

We report performance in terms of percent correct (Fig. 4), but we confirmed that the metric d′ yields the same conclusions (Fig.5 e, Supplementary Fig. S8), and that the effects on performance are not an artifact of bias (Supplementary Analysis 1).

**Statistical tests**

The performance of individual rats in each flanker condition is assumed to be the parameter of a stationary binomial distribution, so finding flanker-caused performance changes amounts to detecting differences in binomial parameters. When computing confidence intervals for differences of binomial parameters, we use the Agresti-Caffo method, in which one modifies a Wald interval by adding two successes and two failures to each estimated proportion(Agresti and Caffo, 2000). Like the Wald interval, the Agresti-Caffo confidence interval uses the Gaussian approximation for binomials, which is not valid if p is near 0 or 1, or n is too small. In our data, $0.55<p<0.85$, and $n>5,000$ for all conditions. We verified that alternative statistics agreed with the significance conclusions of Agresti-Caffo for representative test cases (specifically a permutation test and a Bayesian MCMC method, not shown). The Agresti-Caffo intervals graphed for each subject may exclude the point of zero difference; this can be used to test a hypothesis at $p>0.05$.

To determine if performance differences between conditions were significant at the population level, we used a 2-wayANOVA (`anovan` in Matlab, linear model, type 3 sum squared error). We tested for reliable changes across conditions accounting for the expected variability for each subject. For each subject and condition, we obtained multiple estimates of the performance based on a non-overlapping sample of N trials selected randomly from the testing phase. We used N=200 trials per estimate. For interleaved trials without flankers (F- mix in Supplementary Fig. S3) the number of estimates ranged from 3-13, compared with 78-253 estimates of F+. For comparing flanker conditions (Fig. 4 and 5) the number of independent estimates per flanker condition and subject ranged from 19-62. The resulting distributions were approximately Gaussian (e.g., when comparing flanker conditions typically 26/28 passed Lilliefors' test, close to the 5% failure rate expected on chance). Nevertheless, because distributions cannot be guaranteed to be Gaussian, we also we also report Friedman's test (`friedman` in Matlab), which has lower power but does not assume normality.

Whenever we made multiple comparisons, we used Tukey's honestly significant difference criterion (`multcompare` in Matlab) with a criterion of p<0.01. For example we tested all pairwise differences between the four stimulus conditions. Tukey's criterion for significance is more stringent, to adjust for the fact that making more comparisons increases the probability that one of them will cross significance by chance.  A graph showing the critical values for hypothesis testing are shown in Supplementary Figure S5 where all six pairwise comparisons are reported. Results were not different with p<0.05 criterion. All tests that were significant with the 2-way ANOVA were also significant for Friedman's test except for one comparison (Supplementary Fig. S5e), discussed there.

The 95% confidence intervals for differences in $d'$ or criterion within individual subjects (Fig. 5e,f) were determined using a Monte Carlo Markov Chain.  For each subject and stimulus condition we sampled the posterior distributions of $d'$ constrained by the number of observed hits, misses, correct rejects and false alarms.  The model assumes the hit and false alarm counts are independent binomial distributions and uses a uniform prior over hit and false alarm rates. The $d'$ or criterion posterior was estimated using WinBUGS and software written by Michael Lee(Lee, 2008).  The distribution of $d'$ difference or criterion difference is sampled by taking the difference between independent random draws of two performance conditions.  The confidence interval is the range of this data after removing the 2.5% highest and 2.5% lowest samples.

**SUPPLEMENTAL MATERIALS**

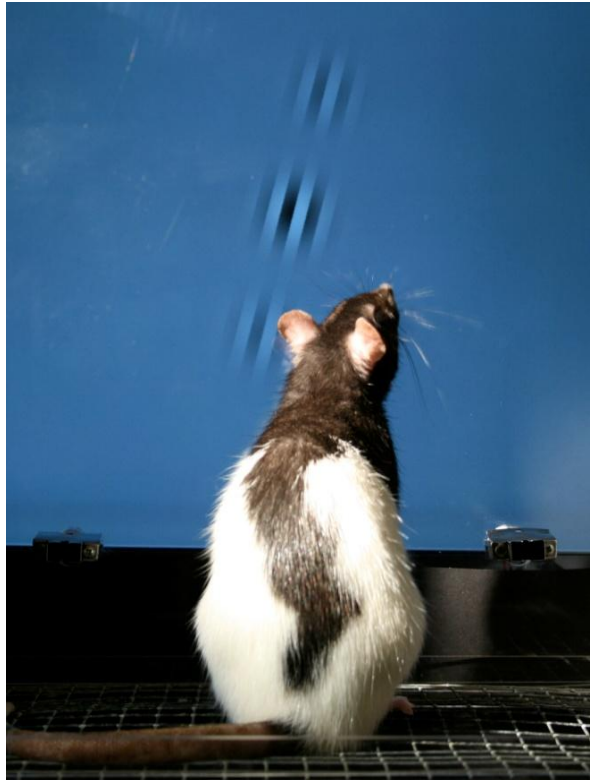

*Fig. S1. Photograph of a rat viewing a target with collinear flankers. The rat is trained to receive water from the port on the right when the central target is present, and the port on the left if it's absent.*

**Supplementary Data 1: Flanking stimuli impair target detection performance**

      The final step in our shaping sequence is the addition of flankers to the detection task (Fig. S2b). This made the task more difficult for rats, presumably for both cognitive and perceptual reasons. This study was not designed to explain this effect, which will be a constant across our compared conditions.

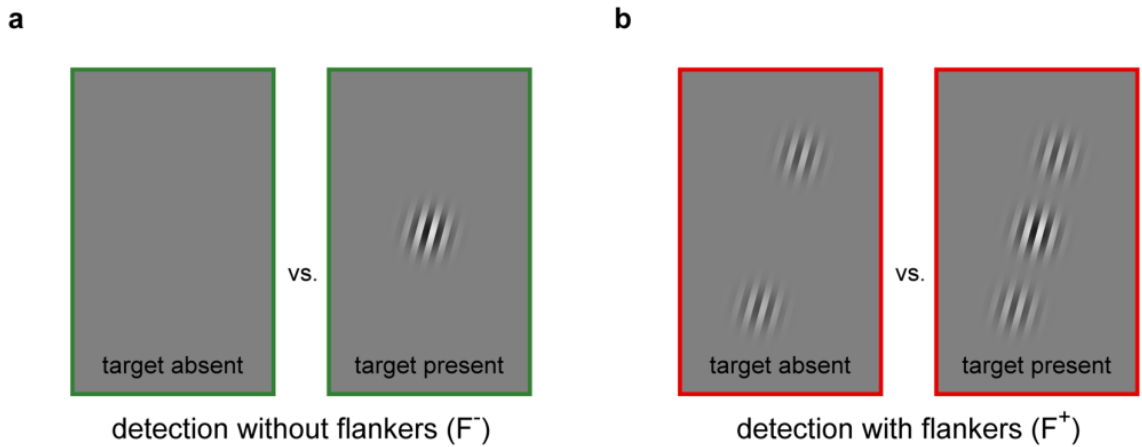

*Fig. S2. Stimuli used for target detection, with or without flankers.*   a) During training (e.g. step 8), the rats perform a detection task when no flankers are present ($F^-$). When the rat initiates a trial, a target appears 50% of the time. When it is present, it is always located in the center of the screen. A rat would receive a water reward by correctly choosing to lick a sensor on the left side if the target were absent, or on the right side if the target were present. b) During the testing phase flankers are present ($F^+$) on 95% of the trials. The flanker's presence, location and orientation carry no information about the correct response. Only one flanker configuration is shown here, though all types were presented during testing (for more types see Fig. 4). In the first analysis all flanker conditions are grouped together.

To quantify the overall influence of flankers on rats' ability to detect a localized target, we compare detection with and without flankers, pooling over all other stimulus parameters ($\theta_T$, $\theta_F$, $\omega$, $S_T$, $S_F$, see Fig. 4). During the continuous block of trials without flankers (Fig. S2a), rats performed about 75% of trials correctly. A single rat's performance is shown in Figure S3a, and for all seven rats in Figure S3b. Flankers impaired performance relative to both the previous block without flankers and the 5% of randomly interleaved trials in the testing step that omitted flankers. We note that performance on trials without flankers was lower when trials with flankers were interleaved as opposed to blocked, although the stimuli were identical. This implies that rats' visual processing or decision strategy depends on the distribution of stimuli over recent trials. For this and other reasons, all other results we show (except Supplementary Fig. S4c) compare only conditions randomly interleaved in the same block.

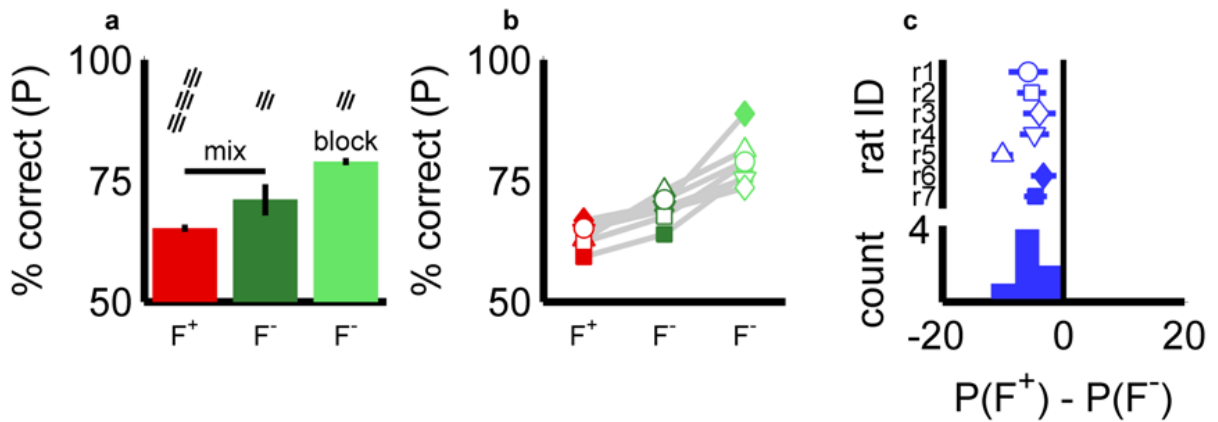

*Fig. S3. Rats are worse at detecting the target when the flankers are present.* a) The percent correct performance (P) of a single rat on three conditions with identical target properties. Error bars indicate 95% binomial confidence intervals. Performance from a block of randomly interleaved trials with flankers ($F^+$) and without flankers ($F^-$) is shown in red and dark green respectively. Light green indicates performance from a continuous block of trials without flankers ($F^-$) before flanking stimuli were introduced. b) Performance of seven rats in the same task, symbols colored as in (b), lines connect symbols representing a single subject (r1-r7). c) The effect size, measured as the difference in percent correct for interleaved trials with and without flankers: $P(F^+) - P(F^-)$. Symbols left of the zero line indicate that percent correct is lower when flankers are present. Effect is shown for each subject (symbol location indicates mean difference, horizontal lines indicate Agresti-Caffo 95% confidence interval, see Methods). For all individual subjects, the deficit with flankers is statistically significant. The effect is also significant if the difference between conditions is assessed at the population level ($p<0.01$, Tukey-Kramer on 2-way ANOVA; $p<0.01$, Tukey-Kramer on Friedman's test, see Methods).

The effect of flankers is summarized as the difference in performance on interleaved trials with and without flankers (Fig. S3c). If detection were independent of stimuli outside the target location, the difference would be zero. These data, however, show the difference to be less than zero, indicating worse performance with flankers.

We note that the addition of flankers also increases the probability that rats respond yes (Fig. S8a). This reflects a large increase in false alarms, and a small increase in the hit rate. We have not attempted to study or interpret this effect. These data are consistent with a modest reduction in "yes" responses due to contrast normalization, together with a large increase in "yes" responses due to task confusion or strategy, but our data by no means establish this interpretation.

We suspect that flankers confuse rats and also exert contrast normalization. Additional experiments would be required to isolate these components**.** Therefore we draw no conclusions about the underlying cause of the net effect of adding flankers, but see (Meier P, 2010). For this study, the effect is a constant across the compared conditions, and performance with flankers is sufficiently high that differences in performance between flanker configuration can be detected.

**Supplementary Data 2. Choosing stimulus parameters**

In order to test for influence of the surround, a detection task must be difficult enough to resolve a difference between conditions, but not so difficult that rats fall to chance performance. Difficulty is influenced by the target's size, contrast and spatial frequency as well as the proximity and contrast of the flanker. We trained rats on easy stimuli, and then varied one or two parameters at a time in separate experiments. These pilot experiments were done on different subsets of the subjects, as well as some subjects that were not in this study.

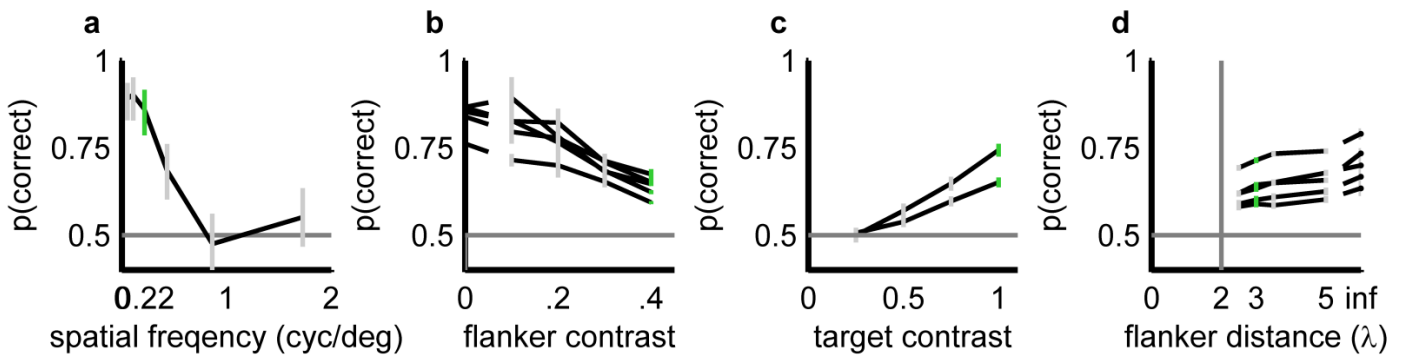

*Fig. S4. Detection performance depends on target spatial frequency, target contrast, flanker contrast and flanker proximity. a) Spatial frequency dependence of a single rat's ability to detect a single large grating patch (Gaussian window of width 21.6 deg/std). Data were collected while randomly interleaving 5 contrasts and 5 spatial frequencies, but only data from 100% contrast (the target contrast used in testing) are shown. Performance was near chance for 0.86 or 1.73 cycles per degree (cyc/deg) at all contrasts, and was near maximum at 0.22cyc/deg at 100% contrast (green), which was used for all other tests in this paper for all rats. Error bars are the 95% binomial confidence interval. b) Detection performance for five rats while learning to detect targets in the presence of flankers of increasing contrast, using 0.22 cyc/deg gratings. Each rat is indicated by a separate curve. As rats demonstrated aptitude on lower flanker contrasts, they automatically graduated to the next higher contrast. All data points involve many hundreds of trials, with the exception of a single data point from one rat at 10% contrast involving only 75 trials. Performance for that rat may appear to be greater than performance with zero contrast flankers, but this difference is not significant. (Agresti-Caffo 95% confidence interval) All other tests in this paper used 40% flanker contrast (green). c) Detection performance from two rats with five randomly interleaved target contrasts, while flanker contrast remained at 40%. All other tests used a target contrast of 100% (green). d) Detection performance from five rats on stimuli with four randomly interleaved flanker distances. The line at $\lambda=2$ represents the distance that full contrast target and flankers begin to perceptually overlap for humans. At $\lambda=5$ both flankers are still on the screen, but are close to the edges. The data at flanker distance of infinity indicates trials where there was no flanker on the screen. All other data in this paper uses $\lambda=3$ (green).*

We tested spatial frequencies spanning 5 octaves (0.05-0.86 cyc/deg) and five contrasts (12.5-100%), for a total of 25 interleaved conditions. This test was performed using a large target with no flankers present (see Fig. 1d, step 7 for example stimulus). At 100% contrast, rats performed sufficiently well on 0.22 cyc/deg (green, Fig. S3a), but not well enough for 0.43 or any higher spatial frequencies. This is consistent with the visual contrast sensitivity previously reported for Long Evans rats; 0.22 cyc/deg is well within their capacity, but a little higher than their peak sensitivity. Gratings were 0.22 cyc/deg for all stimuli presented during the testing step in the main results, as well as all of the other sub-panels in Figure S4.

When flankers were present, they induced a deficit in pe*rf*ormance that increased with their contrast (Fig. S4c). This sub panel is the only data in this paper (other than Fig. S3a,b) that was not randomly interleaved between conditions, but was derived from training steps in which the rats were automatically shaped to higher contrast flankers (see Fig 1d, step 9 for example stimulus). We know that the rat's behavior during this learning stage was not constant over time. In principle, learning effects could have increased a rat's performance at higher contrasts which were tested at a later date. Yet flanker impairment overwhelms any such learning improvements. We include this panel because it is informative about the rats learning and about the difficulty of the detection task in the presence of flankers. The trend is the same for other rats in which a range of flanker contrasts were revisited after learning (data not shown).

After rats were well trained on detecting targets in the presence of flankers, we examined the influence of target contrast. Interestingly, rats were sensitive to the contrast of the target across the entire range of our linearized monitor. If stimuli are supra-threshold on a detection task, one would expect performance to plateau. However, if increasing contrast improves performance, this is consistent with the target stimulus being at or near threshold for detection. Since contrast limited performance, even at the upper end of our display, we chose a high contrast target, so that we could better resolved differences in performance (near 75% correct). We believe that the threshold for detection is shifted to a higher contrast by virtue of the flankers in the surround.

Flankers have a greater influence on target detection when they are closer. We note the possibility that subjects would be best when flankers are at a distance of $3\lambda$, which they had prior training with. However, we observer rats perform worse at $2.5\lambda$ and better at $3.5\lambda$. Despite the stronger effect at $2.5\lambda$, our main results use a flanker distance of $3\lambda$ because previous work in human psychophysics and cat V1 neurophysiology find effects at this distance. Additionally, the tails of the Gaussian masks for target and flanker begin to overlap and $2.5\lambda$, even though we truncate tails at 4 std. There is no overlap for our flankers and targets at a distance of $3\lambda$.

Taken together these data suggest that the flanker stimuli used in our study are perceptually difficult for rats, because the spatial frequency is high for a rat, the grating patch small, and the flanker contrast high. The sensitivity to changes in contrast of the target or the contrast or distance of the flankers is all consistent with expectations from contrast normalization. Due to confounding cognitive factors we do not purport to measure the strength of contrast normalization from these data.

**Supplementary Data 3: Statistics for all pairwise comparisons of flanker configurations**

In the main paper we show statistics for comparisons between the collinear condition and two others: the condition that differs only in flanker orientation (Fig. 4c) and the condition that differs only in the flankers' angular position (Fig 4d).

For the four stimulus conditions tested, there are a total of six pair-wise comparisons. Rats performed worse on collinear stimuli than on any of the other three stimulus conditions (Fig. S5a,c,e, left column). Each of these comparisons is the result of disrupting collinearity in a different way: by changing the flanker orientation (a, same as Fig. 4c), by changing the flankers' angular position (c, same as Fig. 4d), or by changing the target orientation (e). Taken alone, it seems possible that the visual system may be sensitive to each of these changes in isolation. However, the other three conditions provide a control (Fig. S5b,d,f; right column); in each the geometry of the stimulus was held constant except for the flanker orientation (b), flanker location (d), or target orientation (f). No individual rats are significant in these three control comparisons (Agresti-Caffo 95% confidence interval), and they do not differ at the population level (Tukey's on ANOVA, $p>0.05$; Tukey's on Friedman's test, $p>0.05$).

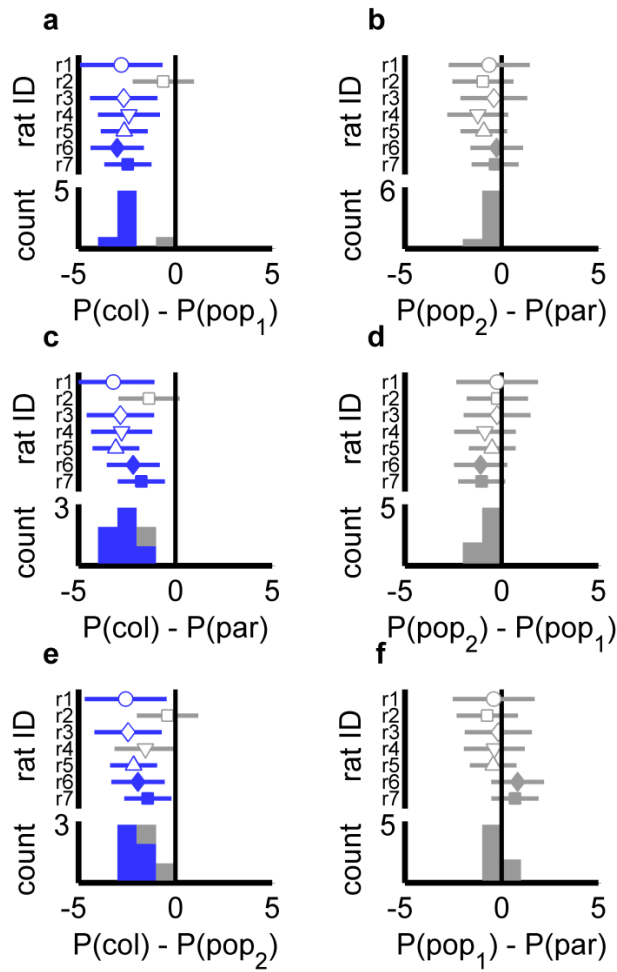

*Fig. S5. Additional pair-wise comparisons between the flanker conditions.* We present all six comparisons between the four stimulus conditions measured by the difference in percent correct. Panels a and c are the same as Figure 4c,d; they are reproduced here to facilitate comparisons. The vertical midline indicates that the performance on the two conditions is not different. Each horizontal line represents the 95% confidence interval for a single subject's data on a given comparison (Agresti-Caffo). The top row of sub panels isolates comparisons between pairs of conditions in which the location of the flanker differs, but other parameters were held constant: a) collinear – parallel and b) popout$_2$ – popout$_1$. The middle row isolates comparisons in which only flanker orientation differs: c) collinear – popout$_1$ and d) popout$_2$-parallel. The bottom row captures comparisons in which only target orientation differs: e) collinear –popout$_2$ and f) parallel – popout$_1$. The three comparisons that contain subjects with significant effects(Agresti-Caffo 95% confidence interval) are best explained by the reduction of performance for collinear stimuli which affects all the comparisons in the left column. The comparisons in the right column are not significant for the rats individually, nor at the population level (see Fig. S6).

To communicate significant differences at the population level, we present the marginal mean performance for all rats (Fig. S6). Each error bar represents the half with of the critical value from Tukey's test of Honestly Significant Difference, which accounts for multiple comparisons. If the error bars from two conditions do not overlap vertically, then the difference between conditions is significant. The figure shows significance for $p<0.05$; we also assessed significance at $p<0.01$. All multiple comparisons in this paper use this same method; here we provide a graphical view to facilitate understanding. Figure S6a summarizes the results for the ANOVA, and Figure S6b summarizes Freidman's test.

In one instance (collinear vs. popout$_2$, Fig. S6), the difference is significant at the population level by 2-way ANOVA but not by Friedman's test after adjusting for multiple comparisons. (Both tests on this condition are significant at $p<0.01$ before adjusting for multiple comparisons). Friedman's test is non-parametric and only acts on the rank-ordered average performance. With only seven subjects, it is quite conservative. It is actually surprising that two of the six comparisons are significant by this test even after adjusting for multiple comparisons. We interpret the collinear condition to be harder for rats than the popout$_2$ condition, based on the results of the 2-way ANOVA and the facts that this was true for 7 of 7 rats, and significant for 5 of 7 rats.

We find it somewhat surprising that there is no discernable effect of popout with respect to the parallel condition (Fig. S5b,f). One might have expected popout conditions to be easier. Maybe the presence of only two flankers does not indicate the uniqueness of the center target as much as a field of distracters would(Nothdurft, 1991; Smith et al., 2007). Alternatively, the orientation difference between target and surround may be important. We used 30° (open symbols in Fig. S3) and 45° (filled symbols), and did not observe a difference. A popout effect might be found at more orthogonal angles(Nothdurft, 1991; Schwartz et al., 2009).

In summary, our only significant differences involved comparisons to the collinear condition. Thus, impairment on the collinear stimulus is the primary effect of pattern-specific processing that we observed.

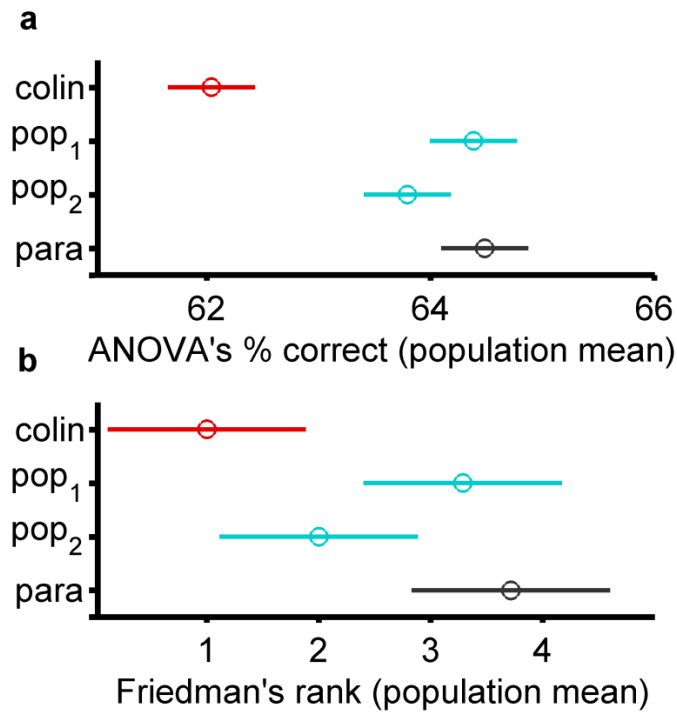

*Fig. S6. Multiple comparison adjustments.* a) A visual summary the 2-way ANOVA after having adjusted for multiple comparisons. The x-axis is the marginal means of the percent correct performance from a population of seven rats. The error bars represent the half width of the critical value for multiple comparisons using Tukey's Honestly Significant Difference at $p<0.05$. If the tails of two error bars do not overlap each other on the vertical axis, then the difference those two conditions are significantly different at $p<0.05$. b) A visual summary the Friedman's test after having adjusted for multiple comparisons using the same method. Friedman's is a non-parametric test that does not assume Gaussianity; it tests for a shift in the location of a probability distribution by analyzing the relative rank of performance across rats and conditions. Friedman's has less power to reveal an effect, especially with the small $N=7$. The collinear condition is reliably the $1^{st}$ rank (worst performance) for all rats. The popout$_1$ and parallel conditions have higher performance. Interestingly, popout$_2$ is reliably the $2^{nd}$ rank, but this test could not resolve a significant difference from the collinear condition. We do not interpret this strongly because the absence of observing an effect is inconclusive, and the population marginal means shown in a), suggests more strongly that popout$_2$ groups with popout$_1$ and parallel more than with collinear. Additionally, the significance of the each of the rats individually (Fig. S5) supports the conclusions of the 2-way ANOVA. All tests that are significant in this figure (both a and b) are also significant at $p<0.01$ (not shown).

**Supplementary Data 4: No influence of phase**

In our experiment, for each flanker orientation and position we also varied sign of the grating ($S_F = \pm1$) corresponding to inverting the luminance of the grating such that dark bars are switched for light bars. Changing the sign is equivalent to a $\pi$ shift in the spatial phase of the grating. Neither the absolute sign of the target or flanker nor the relative sign between them had an effect on performance. Here we show there is no difference between the phase aligned and phase reversed collinear stimuli using the same data from the seven rats reported in the main paper (Fig. S7a,c). This experiment used one spatial frequency (0.22 cyc/deg), one flanker distance (3λ, 13.6°), and two relative phases (0, π). Our data do not exclude the possibility that the rat visual system is sensitive to phase differences with other choices of these parameters.

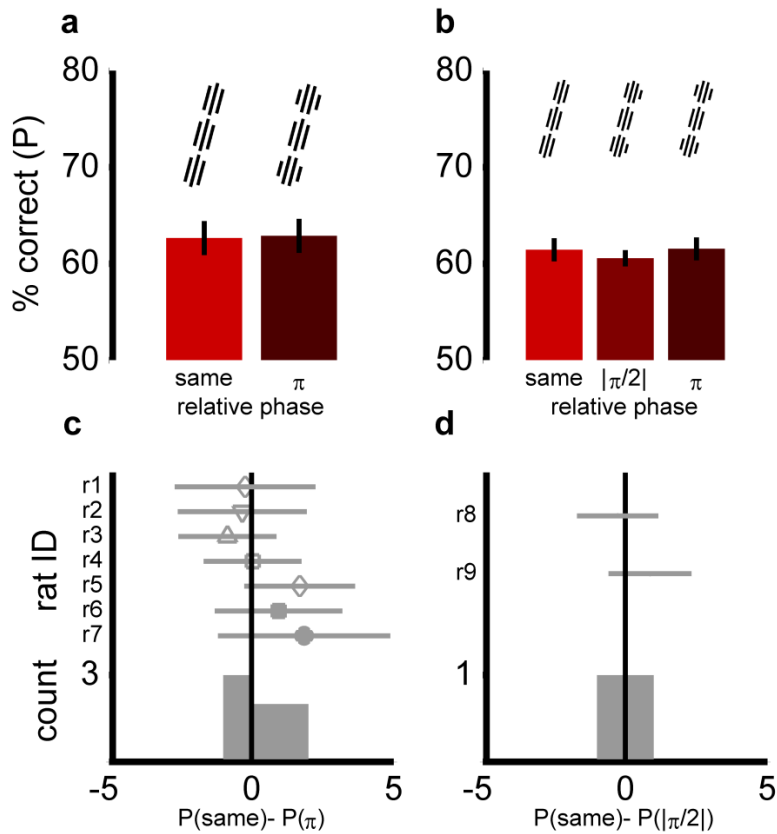

Fig. S7. Phase of grating has no influence on detection with flankers. a) A single rat's performance on collinear stimuli that were either phase-aligned or shifted by pi (sign-reversed). This is the same data as Fig. 4, reanalyzed to show that there is no effect of phase. b) A single rat in a separate pilot study in which four target phases and four flanker phases were used. Conditions are grouped by the phase difference between target and flanker. c) The difference in percent correct between phase-aligned and phase-reversed trials for the seven rats in the main study. No individual rats show significant differences. The population as a whole shows no indication that the relative phase of the flankers influences the rats detection performance ($p>0.05$, ANOVA; $p>0.05$, Friedman's test). d) The same is true for the pilot study with more phases. All three pair-wise comparisons are insignificant for each individual rat ($p<0.05$, Agresti-Caffo interval). Shown is the difference between phase aligned and stimuli in which the phase was shift ±π/2. No population statistic was performed because N=2.

In a pilot study (N=2) we tested four different absolute grating phases for target and flanker, using the same class of flanker stimuli as this study, but with some differences: the target contrast was weaker (60% instead of 100%) and rat performed a two alternative force choice between two simultaneously presented stimuli, instead of a single stimulus yes-no task. The relative phase between target and flanker did not influence the detection of the target (Fig. S7b,d).

Since there is no evidence in our data of sensitivity to sign or phase, we have no reason to include it as a relevant parameter in the schematic model presented in the main paper (Fig. 6). The model is sufficient to explain all of the results in this paper, without adding a term in the surround for the relative phase.

The influence of oriented contrast in the surround is not phase sensitive in many previous psychophysical(Xing and Heeger, 2001) and physiological experiments(Webb et al., 2005). However, sometimes the relative phase of the surround does play a role(Ejima and Takahashi, 1985; Williams and Hess, 1998). For primates performing contour detection tasks, spatial pattern summation is phase insensitive in the fovea, but in the periphery it is not(Chen and Tyler, 1999). While phase sensitivity attests to the specificity of surround processing, it has been argued that invariance to phase is a useful aspect of surround processing, and such invariance continues to be applied in computational models of contour integration(Hansen and Neumann, 2008).

**Supplemental Experimental Procedures 1: Water restriction, training schedule and environment**

Beginning around post-natal day 30, rats were restricted from free access to water, instead working for water reward in the training environment. Water was earned solely by correctly performing tasks ("closed economy" training) as long as animals maintained adequate health and hydration. This was assessed by daily weight and health inspections. Supplemental water (or hydrating snacks, e.g. carrot slices) were provided as needed, for less than 25% of rats during initial shaping, and rarely after 2 weeks. On most days, including weekends, animals were transferred to the training chamber where they could freely perform trials for around 90 minutes. On days where sessions were skipped, rats were provided with about an hour of free access to water, but we avoided this because it reduced their motivation to perform trials on the following day. Continuous access to rat chow was provided both in their home cages and the training chamber to stimulate desire for hydration. We phase-reversed the light cycle in the room so that rats trained during the day in dark rooms illuminated almost exclusively by the glow of 7 computer displays. When overhead lighting was required for working in the room during daytime, we typically filtered out the high wavelengths visible to rats(Jacobs et al., 2001) and worked in red light (Encapsulite Intl., red filter, 625nm cutoff, 48SOR20T12).

**Supplemental Experimental Procedures 2: Training system**

The system consists of an array of stations, each controlled by a computer running Windows XP-Pro and Matlab. An additional computer is used as a single point of control for all the stations (via custom TCP/IP code in Java/Matlab). This allows sessions to be started/stopped en masse, management of subject information and trial records in coordination with a database (Oracle 10g Express Edition), and centralized management of individualized training sequences and parameters for each subject.

The system design facilitates either live-in training or easy swap-in of groups of animals. The animals in this study were members of one of 6 groups that were swapped in daily. Since rats can earn hydration adequate for health in roughly one hour of trials, we found it efficient to limit each animal's daily access to trials to sessions of about this duration. This focuses their motivation for performing trials correctly throughout the session and multiplies the number of subjects that can be trained using a given amount of available hardware.

Each training box is 35 cm wide x 18 cm deep x 30 cm high, with detector/reward ports positioned along the front wall spaced 9 cm apart 6 cm above the floor grating. The detector housings protrude 2.5 cm into the box, and the CRT is positioned 5 cm behind their wall. At the center port, a rat's eye is roughly 10 cm from the monitor, 10 cm below its center. At this position, the monitor display subtends 104° of visual angle, and the target grating at the CRT center is roughly 14 cm away and spans 6.8° per std of the Gaussian mask. This is closer than the distance established for maximum behavioral visual acuity in rats (20-30 cm; Wiesenfeld, Branchek 1976).

Accurate reward delivery requires a system that does not store pressure in unexpected ways. At such small volumes and flow times, tiny differences in port geometry likely affect the amount of reward and its accessibility due to being wicked away or the force with which it is ejected, and reward volumes/utility may not be linear with respect to open valve duration. This can cause strong side biases. We monitored behavioral trends for side bias and found we could control it using timeouts and correction trials alone; we only sporadically and imprecisely verified that left and right reward drops looked roughly similar in size. We have since added a syringe-pump based reward system that is more accurate and requires less maintenance (New Era Pump Systems, Inc. NE-500).

We used the parallel port for electronic interface of the valves and sensors with the computer, using PortTalk (Craig Peacock, http://www.beyondlogic.org/porttalk/porttalk.htm) via an open source Matlab wrapper (http://psychtoolbox.org/wikka.php?wakka=FaqTTLTrigger).

The sound for requests is a tone with energy at each octave spanning the frequency range of the speakers (some range within 20-20,000 Hz); for correct responses, a tone with the same harmonic structure is played a perfect fourth above the request tone, creating a harmonic resolution. The sound for incorrect responses is a chord made from two tones of the same harmonic structure separated by a tritone, the maximally dissonant interval. The sound for inappropriate responses is broadband noise. Sometimes a drop of water lodged in a port is sufficient to break the infrared detector beam, in which case, the corresponding sound plays continuously until the water is cleared by the rat (after some experience in the box, rats are observed to clear ports in this way even during a period of no trial activity, indicating a possible preference for silence to the continuous sounds). No provisions were made for sound isolation. Sounds from adjacent boxes are quite audible in any given box, but quite attenuated relative to that box's local sounds. The CRTs for adjacent boxes are not visible when operating the ports because stations are separated by opaque dividers.

Our software architecture facilitates the design of arbitrary task structures and manages each subjects' progress through their own training protocol. A *protocol* specifies a sequence of *training steps*, each with a *trial manager* (defining task structure – e.g. two-alternative-forced-choice), *stimulus manager* (defining audiovisual stimuli and their parameters), *reinforcement manager* (specifying reward/timeout reinforcement rules), and *graduation criteria* (regulating progression through steps).

**Supplemental Experimental Procedures 3: Stimulus display**

We used CRT displays (NEC FE992, 19") because of their fast refresh and phosphor decay times; CRTs are generally better suited for visual psychophysics than LCD flat panels because their timing artifacts and brightness artifacts are more consistent and better understood. The CRT linearization table was created by fitting a power law with gain and offset ($y=b*x^{\gamma}+m$) to photodiode measurements (Thorlabs, PDA55), and then computing the inverse function ($x=[(y-m)/b]^{1/\gamma}$) for each RGB channel independently. Each value was measured as the average height of the smoothed peak of the phosphor decay curve recorded with that value presented in a rectangle occupying the central 60% of screen for 9 frames at 100 Hz and 1024x768 resolution. Gun values were measured in increasing order rather than randomized. The resulting tables were then verified to have linearized grayscale output to within 0.5% using the same method.

**Table S1: Training Details**

| Step | Description | Visual Stimulus | Graduation Criterion | Goal | Duration (trials ± std) | Duration (days ± std) |
|---|---|---|---|---|---|---|
| 1 | Free drinks | none (mean gray) | 4 trials in 1 min | get water from port | 100 ± 65 | 1 ± 0 |
| 2 | Earned drinks | none (mean gray) | 5 trials/min for 2 min | alternate responses | 170 ± 100 | 2 ± 2 |
| 3 | Faster drinks | none (mean gray) | 6 trials/min for 3 min | sustain interest | 170 ± 100 | 2 ± 2 |
| 4 | Side rewards | big grating | 5 trials/min for 5 min | 2AFC trial structure | 350 ± 220 | 8 ± 5 |
| 5 | Easy detection | same big grating | 85% correct | 1st visual learning | 4,100 ± 3,300 | 17 ± 13 |
| 6 | Linearized luminance | lower contrast | 85% correct | new stim | 1,700 ± 1,300 | 9 ± 9 |
| 7 | Thinner | double target cyc/deg | 85% correct | new stim | 900 ± 1,100 | 4 ± 3 |
| 8 | Smaller | target mask 1/3rd size | 85% correct | new stim | 7,300 ± 8,200 | 35 ± 45 |
| 9 | Add flanks | slowly raise flank contrast | 80% correct | new stim | 11,000 ± 9,000 | 25 ± 13 |
| Test | Flankers | See Fig. 4b | none | collect data | 55,000 ± 23,000 | 98 ± 36 |

*Table 1. Shaping sequence and training steps. Step numbers correspond to the numbers in the colored bars in the Figure 1c training timeline. All rats progressed from step 1 to 9 in increasing order. Two rats (r1, r2) performed contrast and spatial frequency varying psychometric curves before step 8 (not listed in this chart, results in Fig. S4a). Another two rats (r4, r8) performed the flanker task with varying target contrast (not listed in this chart, Fig. S4c). Other supplementary tests either used data from the training sequence (flanker contrast, step 9, Fig. S4b), the main testing step (influence of luminance sign, Fig. S7a,c), or after the main testing step (flanker distance, Fig. S4d).*

Since the background grey level was set equal to the mean luminance of the grating, linearizing has the consequence that a blurry optics system with spatial resolution lower than the spatial frequency of the gratings will not perceive a luminance difference at any point in the Gaussian patch of the grating, but only a smooth grey at the same luminance as the background.

We confirmed that this was the case for sufficiently high spatial frequencies for human observers. The linearization range was chosen such that the gratings were effectively lower contrast than in previous steps and the mean grating luminance/background grey was brighter than in previous steps. The minimum, mean, and maxiumum luminance were set to 4, 42 and 80 cd/m$^2$, respectively (Colorvision, spyder2express).

**Supplementary Analysis 1: Considering signal detection theory**

In the main paper we summarize performance as percent correct, but using d′ does not alter the character of the results (Fig. S8). Signal detection theory measures attempt to unconfound bias and sensitivity, both of which affect percent correct. The metric of percent correct can be misleading when comparing results of subjects with very different overall biases. In this study we do not directly compare any rat's absolute performance to another rat's. Rather we test within subject for significant differences between stimulus conditions that are randomly interleaved. When the hit or false alarm rate is near 0 or 1, it is possible that the metrics of d′ and percent correct could give different trends or statistical significance when comparing conditions. In our data, the hit and false alarm rates are bounded by [0.15 0.85] for all rats and conditions. Thus our data do not occupy the extreme region where one would expect significant differences between the metrics.

One advantage of using percent correct, when the data permit, is that it avoids the need to make assumptions which we cannot verify.Specifically, signal detection theory separates bias from sensitivity by assuming normality and homoscedacity of the stimuli distributions being distinguished – d′ summarizes their discriminabilty as the separation of their means in units of their common standard deviation. Another reason we prefer percent correct is that, unlike d′, it monotonically increases with reward; thus, tracking what rats naturally prioritize. We report the assumption-free data on hit rate and false alarm rate from which d′ and other metrics of signal detection theory are easily calculated.

Nevertheless comparisons based on d′ are provided for comparison (Fig. S8). Across every major comparison in this study, the two metrics agree in individual significance, population trends, and relative magnitude. Specifically, we find that both measures indicate all flanker types impoverish performance (Fig. S6b,c), collinear flankers impair performance more than parallel flankers (Fig. S6 e,f), and collinear flankers maintain this impairment regardless of the relative contrast sign between target and flanker (Fig. S6 h,i).

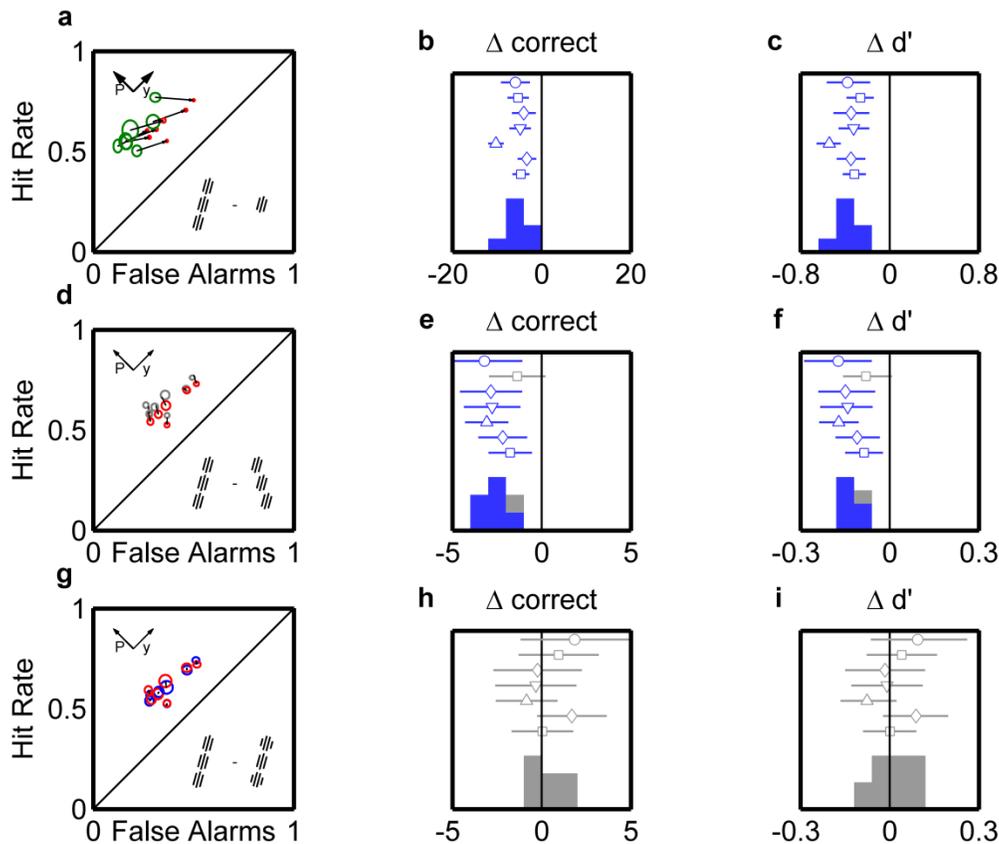

*Fig. S8. Percent correct and d' yield the same conclusions.* The raw data are presented for three experiments: the influence of flankers (a,b,c), the influence of collinearity (d,e,f), and the influence of phase (h,i,j). The purpose is to validate that the conclusions are the same using percent correct or d' as a performance metric. a) Same data as in Figure S3, comparing the difference between two conditions summarized by the icons lower right of the panel. The ellipses indicate performance of each rat on interleaved trials with the flanker present (red) or absent (green). Arrows indicates a single subject's change in hit rate and false alarm rate. The axis of hit rate and false alarm can be rotated to indicate percent correct performance (P) and percent yes responses (Y), as indicated by the inset axis in the upper left. All rats have a decrease in the percent correct when flankers are present. b) Performance reduction quantified by percent correct using Agresti-Caffo 95% confidence interval(identical to Fig. S3c). Each horizontal line is the confidence interval for a single rat. The histogram of the mean change in performance color coded blue for each rat that is individually significant, and grey otherwise. c) Performance reduction quantified for d' using the 95% confidence interval of an MCMC simulation. Panels (d,e,f) employ the same convention as (a,b,c), except comparing collinear flanker condition (red) to the parallel flanker condition (gray). Data are the same as Figure 5. e) difference by percent correct metric (identical to Fig. 5d). f) difference by d' metric (identical to Fig. 5e). Panels (h,i,j) show the lack of an observed difference between collinear stimuli that where phase aligned (red) or phase reversed (blue). The data are the same as Figure S7 a,c. The effects in panels b, c, e and f were all significant by ANOVA ($p<0.01$ and by Friedman's test ($p<0.01$), adjusted for multiple comparisons where required. There was no significant difference in phase (panels h and i) at the population level ($p>0.5$ both ANOVA or Friedman's test).